\documentclass{ws-procs9x6}

\begin{document}

\title{Missing Resonances in Kaon Photoproduction on the Nucleon}

\author{T. Mart, A. Sulaksono}

\address{Departemen Fisika, FMIPA, Universitas Indonesia, Depok 16424, Indonesia}

\author{C. Bennhold}

\address{Center for Nuclear Studies, Department of Physics, The George 
         Washington University, Washington, D.C. 20052, USA}  

\maketitle

\abstracts{New kaon photoproduction data on a proton, $\gamma + p \to K^+ + \Lambda$, 
  are analyzed using a multipole approach. The background terms are given
  in terms of gauge invariant, crossing symmetric, Born diagrams with
  hadronic form factors, while the resonances are parameterized using Breit-Wigner 
  forms. Preliminary results suggest a number of new resonances, as 
  predicted by many quark model studies.  
  A comparison between the extracted multipoles and those obtained from
  KAON-MAID is presented.}

\section{Introduction}
Considerable theoretical and experimental efforts to understand
the structure of the nucleon have been devoted for more than fifty years. 
A consequence of this substructure is the nucleon 
resonance spectra found in the region between 1 - 3 GeV. This region is 
accessible neither through perturbative QCD
at high energies nor through chiral perturbation theory at low energies. 
There is hope that lattice QCD will provide answers through numerical techniques.
Meanwhile, our knowledge of resonance physics
comes mostly from phenomenology. This paper presents
a new investigation of the nucleon spectrum through the photoproduction of a kaon
on a proton. To this end, we will make use of a multipole approach to describe
the resonant states. Such studies have been reported by previous 
authors\cite{thom1966,Tanabe:zu}, albeit using a limited data base. 
We report preliminary results using the new data that have become available 
this year from SAPHIR\cite{Glander:2003jw} and JLAB\cite{McNabb:2003nf}.

\section{Formalism}
\subsection{Background Amplitudes}
The basic background amplitudes are obtained from a series of tree-level Feynman 
diagrams, shown in previous work\cite{thom1966,Lee:1999kd,Adelseck:ch,David:1995pi}.
They contain the standard 
$s$, $u$, and $t$-channel Born terms along with the $K^*(892)$ and $K_1(1270)$ 
$t$-channel poles. Apart from the $K_1(1270)$ exchange, these background terms are
similar to the ones used by Thom\cite{thom1966}. The importance of the 
$K_1(1270)$ intermediate state has been pointed out in Ref.\cite{Adelseck:yv}. 
To account for the hadronic structures of interacting 
baryons and mesons we include the appropriate hadronic form factors 
in the hadronic vertices by utilizing Haberzettl's 
method\cite{Haberzettl:1998eq} in order 
to maintain gauge invariance of the amplitudes. Furthermore, to comply 
with the \textit{crossing symmetry} requirement we use a special
form factor in the {\it gauge terms},
as has been proposed by Davidson and Workman\cite{Davidson:2001rk}.
Thus, compared to the previous pioneering work\cite{thom1966}, the
major advancement in the background sector is the use of hadronic 
form factors to control its contribution in a gauge-invariant 
and the crossing-symmetric fashion.

\subsection{Resonance Amplitudes}
The resonant electric and magnetic multipoles for a state with the mass $M_R$, width 
$\Gamma$, and angular momentum $l$ are assumed to have the Breit-Wigner form
\begin{eqnarray}
  \label{eq:e_multipole}
  E_{l\pm},M_{l\pm} &=& \left\{\frac{1}{k_R\, q_R\, j_\gamma(j_\gamma+1)}\,\frac{v_l(qR)}{v_l(q_RR)}
    \right\}^{1/2}\, \frac{M_R\,\Gamma\sqrt{\Gamma_{\gamma E(M)}\Gamma_K}\, e^{i\theta}}{
      M_R^2-s-iM_R\Gamma} ~,~~~~
  \label{eq:m_multipole}
\end{eqnarray}
where $s$ represents the square of the total c.m. energy, $k_R$ and $q_R$ are the
photon and kaon momenta evaluated at the resonance's pole ($s=M_R^2$), 
$\theta$ is the phase angle, $j_\gamma =l\pm 1$ for $E_{l\pm}$ and 
$j_\gamma = l$ for $M_{l\pm}$. 
The factors $\Gamma_{\gamma E(M)}$ and $\Gamma_K$
represent the branching ratios of the resonance into $\gamma p$ and
$K^+\Lambda$, respectively.

\begin{figure}[!t]
\centerline{\epsfxsize=11cm\epsfbox{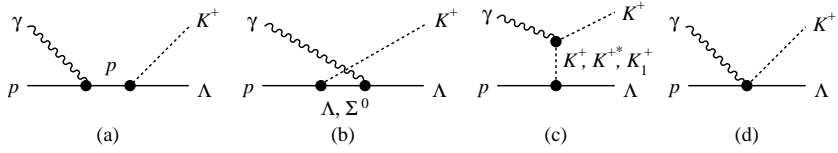}}   
\caption{Contribution of the $s$-, $u$-, and $t$-channel to 
  the background amplitudes for kaon photoproduction on the nucleon
  $\gamma + p \to K^+ + \Lambda$. The contact diagram (d) is
  required to restore gauge invariance after introducing hadronic form factors.
  \label{born}}
\end{figure}

The $v_l(qR)$ denote Blatt-Weisskopf barrier penetration factors 
which accounts for the dependency of partial decay widths on the momentum and 
are given by\cite{blatt-weisskopf}
\begin{eqnarray}
  \label{eq:barrier}
  v_0(x) &=& 1 ~,\nonumber\\
  v_1(x) &=& x^2/(1+x^2) ~,\nonumber\\
  v_2(x) &=& x^4/(9+3x^2+x^4) ~,\\
  v_3(x) &=& x^6/(225+45x^2+6x^4+x^6) ~,\nonumber\\
  v_4(x) &=& x^8/(11025+1575x^2+135x^4+10x^6+x^8)\nonumber ~,
\end{eqnarray}
Following Thom\cite{thom1966}, in this calculation the interaction radius 
$R$ has been fixed at about one Fermi ($1/R=200$ MeV).
All observables can be calculated from the CGLN amplitudes\cite{Knochlein:1995qz}
\begin{eqnarray}
  \label{eq:cgln}
  {F} &=& i \bm{\sigma}\cdot\bm{\epsilon}\, F_1 + 
  \bm{\sigma}\cdot\hat{\bm{q}}\,\bm{\sigma}\cdot (\hat
  {\bm{k}}\times\bm{\epsilon})\, F_2 + i\bm{\sigma}\cdot
  \hat{\bm{k}}\,\hat{\bm{q}}\cdot \bm{\epsilon}\, F_3 +
  i\bm{\sigma}\cdot\hat{\bm{q}}\,\hat{\bm{q}}\cdot \bm{\epsilon}\, F_4 ~,~~~
\end{eqnarray}
where the amplitudes $F_i$ are related to the electric and magnetic multipoles
given in Eq.\,(\ref{eq:e_multipole}) for up to $l=4$ by
\begin{eqnarray}
  \label{eq:f1}
  F_1 &=& E_{0+}-{\textstyle \frac{3}{2}}(E_{2+}+2M_{2+})+
          E_{2-}+3M_{2-}\nonumber\\
      & & +{\textstyle \frac{15}{8}}(E_{4+}+4M_{4+})
      - {\textstyle \frac{3}{2}}(E_{4-}+5M_{4-}) \nonumber\\
      & & +\,3\left\{ E_{1+}+M_{1+}-{\textstyle \frac{5}{2}}(E_{3+}+3M_{3+})
            +E_{3-}+4M_{3-}\right\}\, \cos\theta \nonumber\\
      & & + \,{\textstyle \frac{15}{2}}\left\{ E_{2+}+2M_{2+} -
          {\textstyle \frac{7}{2}}(E_{4+}+4M_{4+}) + 
          E_{4-}+5M_{4-}\right\}\,\cos^2\theta\nonumber\\
      & & +\, {\textstyle \frac{35}{2}}(E_{3+}+3M{3+})\,\cos^3\theta
          + {\textstyle \frac{315}{8}}(E_{4+}+4M_{4+})\,\cos^4\theta ~,\\
  \label{eq:f2}
  F_2 &=& 2M_{1+}+M_{1-}-{\textstyle \frac{3}{2}}(4M_{3+}+3M_{3-})
          \nonumber\\
      & & + 3\left\{3M_{2+}+2M_{2-} - {\textstyle \frac{5}{2}}
          (5M_{4+}+4M_{4-})\right\}\,\cos\theta \nonumber\\
      & & + {\textstyle \frac{15}{2}}(4M_{3+}+3M_{3-})\cos^2\theta
          + {\textstyle \frac{35}{2}}(5M_{4+}+4M_{4-})\cos^3\theta ~,\\
  \label{eq:f3}
  F_3 &=& 3\left\{\, E_{1+}-M_{1-}-{\textstyle \frac{5}{2}}(E_{3+}-M_{3+})+
          E_{3-}+M_{3-}\right\} \nonumber\\
      & & +\, 15\left\{\, E_{2+}-M_{2+}+E_{4-}+M_{4-}-
          {\textstyle \frac{7}{2}}(E_{4+}-M_{4+})\right\}\,\cos\theta\nonumber\\
      & & +\, {\textstyle \frac{105}{2}}(E_{3+}-M_{3+})\,\cos^2\theta + 
          {\textstyle \frac{315}{2}}(E_{4+}-M_{4+})\cos^3\theta ~,
\end{eqnarray}
\begin{eqnarray}
  \label{eq:f4}
  F_4 &=& 3\left\{\, M_{2+}-E_{2+}-M_{2-}-E_{2-}- {\textstyle \frac{5}{2}}
          (M_{4+}-E_{4+}-M_{4-}-E_{4-})\right\}\nonumber\\
      & & + 15 (M_{3+}-E_{3+}-M_{3-}-E_{3-})\,\cos\theta
          + {\textstyle \frac{105}{2}}(M_{4+}-E_{4+}\nonumber\\
      & & -M_{4-}-E_{4-})\,\cos^2\theta ~.
\end{eqnarray}
These amplitudes are combined with the CGLN amplitudes obtained from the background
terms\cite{thom1966,David:1995pi} and substituted into Eq.\,(\ref{eq:cgln}).

\section{Results and Discussion}
In Ref.\cite{Tanabe:zu} Tanabe {\it et al.} put forward a model
which contains the resonances in the partial waves $S_{11}$, $P_{11}$, $P_{13}$,
$D_{13}$, and $G_{17}$ resonances to fit the old $K^+\Lambda$
photoproduction data. It is well known that the $K^+\Lambda$
threshold region is dominated by the states $S_{11}(1650)$, $P_{11}(1710)$, 
and $P_{13}(1720)$. For our analysis, we allow states up to spin 7/2 to contribute.
Sample preliminary results are shown in Table 1, where we explored the sensitivity 
of the data to two separate resonant states at different energies in the same
partial wave, as predicted in some quark models\cite{capstick98}.
In the fitting procedure, we constrain the mass of all 
resonances to vary between 1600 and 2200 MeV. 
For certain partial waves, such as $D_{15}$ and $F_{17}$, the fit generates
spurious states in close proximity to each other (Model I). 
Yet, fits with only one state in
each of these partial waves lead the states with 
masses quite different from what is found before (Model II), indicating that the results
presented here cannot be yet be interpreted as signaling new resonances in these
partial waves.
 
\begin{table}[!t]
\tbl{Resonance states and their parameters extracted
  from fit to models I and model II. 
  The units for $M$, $\Gamma$, $\sqrt{\Gamma_\gamma\Gamma_K}$,
  and $\theta$ are given in MeV, MeV, $10^{-3}$ MeV, and deg., respectively.}
{\footnotesize 
\renewcommand{\arraystretch}{1.1}
\begin{tabular}{|ll|r|r|r|r||r|r|r|r|}
\hline
 &  & \multicolumn{4}{c||}{Model I} & \multicolumn{4}{c|}{Model II}\\
\cline{3-6}\cline{7-10}
\multicolumn{2}{|c|}{Resonance} & $M~$& $\Gamma~$& $\!\!\sqrt{\Gamma_\gamma\Gamma_K}\!\!$ &
$\theta~~~$ & $M$ & $\Gamma~$ & $\!\!\sqrt{\Gamma_\gamma\Gamma_K}\!\!$ & $\!\!\theta~~~$ \\
\hline
$S_{11}$ & $E_{0+}$ & 1610 & 365 & 3.38   & 13.8     & 1610 & 354 & 4.12  &13.5    \\
$S_{11}$ & $E_{0+}$ & 2043 & 283 & 7.76   & $-76.5$  & 2200 & 400 &$-4.94$&91.2    \\
$P_{11}$ & $M_{1-}$ & 1728 & 100 & 6.17   &$-180.0$  & 1707 & 100 &5.00   &152.3   \\
$P_{11}$ & $M_{1-}$ & 1893 & 312 & 12.28  & 87.7     & 1836 & 100 &1.92   &$-180.0$\\
$P_{13}$ & $E_{1+}$ & 1654 & 184 & 5.79   & $-42.6$  & 1688 & 121 &5.45   &$-8.26$ \\
         & $M_{1+}$ & -~~~ & -~~~&$-3.29$ & -~~~     &-~~~  &-~~~ &$-2.97$&  -~~~  \\
$P_{13}$ & $E_{1+}$ & 1909 & 240 &$-3.68$ & 140.3    & 1850 & 400 &3.67   &$-41.1$ \\
         & $M_{1+}$ & -~~~ &-~~~ & $-1.33$& -~~~     & -~~~ & -~~~&$-2.61$&  -~~~  \\
$D_{13}$ & $E_{2-}$ & 2200 & 183 & 4.45   &  $-94.2$ & 1912 & 148 &$-5.72$& 156.7  \\
         & $M_{2-}$ & -~~~ & -~~~&$-1.11$ &  -~~~    &-~~~  &-~~~ & 1.64  &  -~~~  \\
$D_{13}$ & $E_{2-}$ & 1741 & 315 & 1.50   & 90.4     & 1750 & 252 &$-0.62$&$-66.9$ \\
         & $M_{2-}$ & -~~~ & -~~~& 4.49   & -~~~     & -~~~ &-~~~ &$-4.15$& -~~~   \\
$D_{15}$ & $E_{2+}$ & 1780 & 151 & 11.59  & $-21.4$  & 2179 & 400 &$-4.07$& 179.9  \\
         & $M_{2+}$ & -~~~ & -~~~&   0.53 & -~~~     & -~~~ &-~~~ &  1.83 & -~~~   \\
$D_{15}$ & $E_{2+}$ & 1774 & 105 & 9.30   & 148.8    & -~~~ & -~~~& -~~~  & -~~~   \\
         & $M_{2+}$ & -~~~ & -~~~&   3.05 &  -~~~    & -~~~ & -~~~& -~~~  & -~~~   \\
$F_{17}$ & $E_{3+}$ & 1921 & 343 &$-11.36$& 180.0    & 1715 & 100 &$-0.69$& 128.7  \\
         & $M_{3+}$ & -~~~ &-~~~ &$-6.01$ & -~~~     & -~~~ & -~~~&  0.94 & -~~~   \\
$F_{17}$ & $E_{3+}$ & 1900 & 315 & 11.64  & 171.7    & -~~~ & -~~~&  -~~~ & -~~~   \\
         & $M_{3+}$ & -~~~ & -~~~& 5.69   &  -~~~    & -~~~ & -~~~& -~~~  & -~~~   \\
$G_{17}$ & $E_{4-}$ & 1723 & 100 & 0.05   & 116.6    & 1730 & 100 &0.25   & 180.0  \\
         & $M_{4-}$ & -~~~ & -~~~& 0.41   & -~~~     & -~~~ &-~~~ & 0.14  & -~~~   \\
$G_{17}$ & $E_{4-}$ & 2082 & 133 &  0.52  &$-$180.0  & 2011 & 400 &1.59   &$-10.1$ \\
         & $M_{4-}$ & -~~~ & -~~~&$-1.19$ &  -~~~    & -~~~ & -~~~&1.37   & -~~~   \\
\cline{1-2}\cline{3-6}\cline{7-10}
\multicolumn{2}{|l|}{$\chi^2/{\rm d.o.f.}$} &\multicolumn{4}{|c||}{0.85}&\multicolumn{4}{c|}{0.91}\\
\hline
\end{tabular} \label{tab:parameter} }
\end{table}

\begin{figure}[!ht]
\centerline{\epsfxsize=11cm\epsfbox{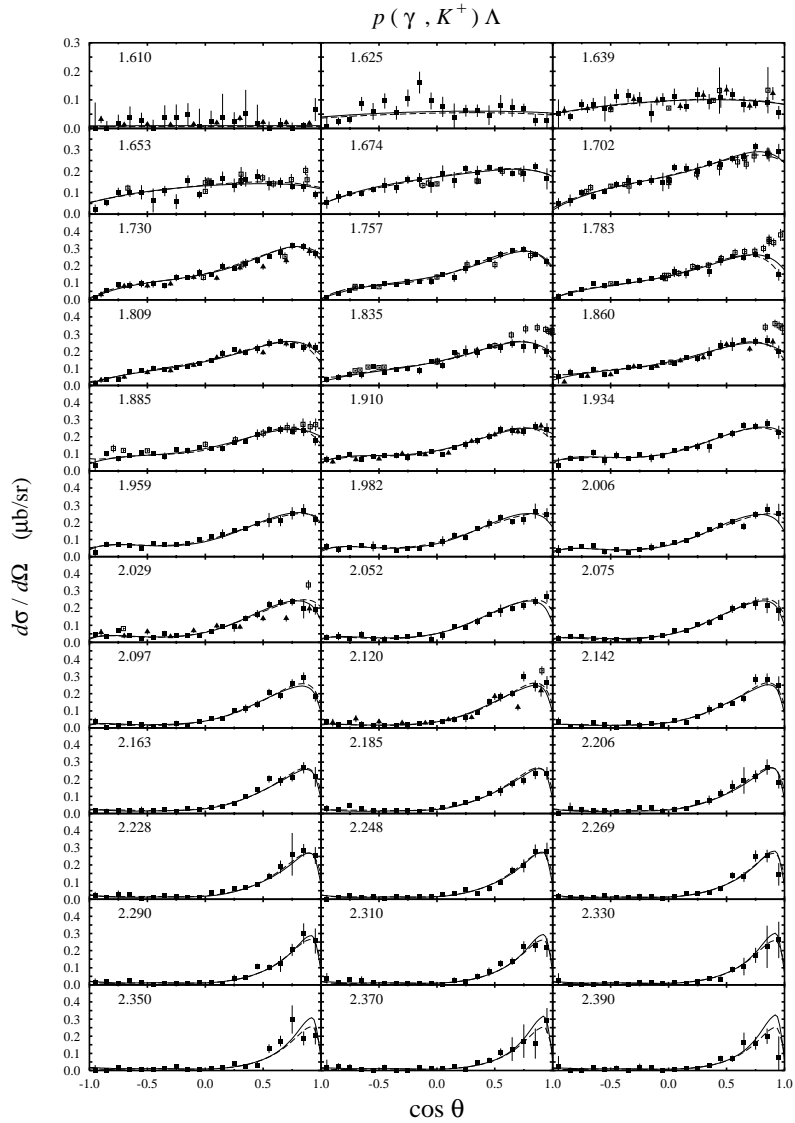}}   
\caption{The $p(\gamma,K^+)\Lambda$ differential cross sections as a function 
  of kaon angle. Dashed and solid lines refer to the Model I and II of 
  Table \protect\ref{tab:parameter}, respectively. Data are taken from
  Ref.\,\protect\cite{Glander:2003jw,Tran:qw}.
  \label{fig:difcs}}
\end{figure}

\begin{figure}[!ht]
\centerline{\epsfxsize=9cm\epsfbox{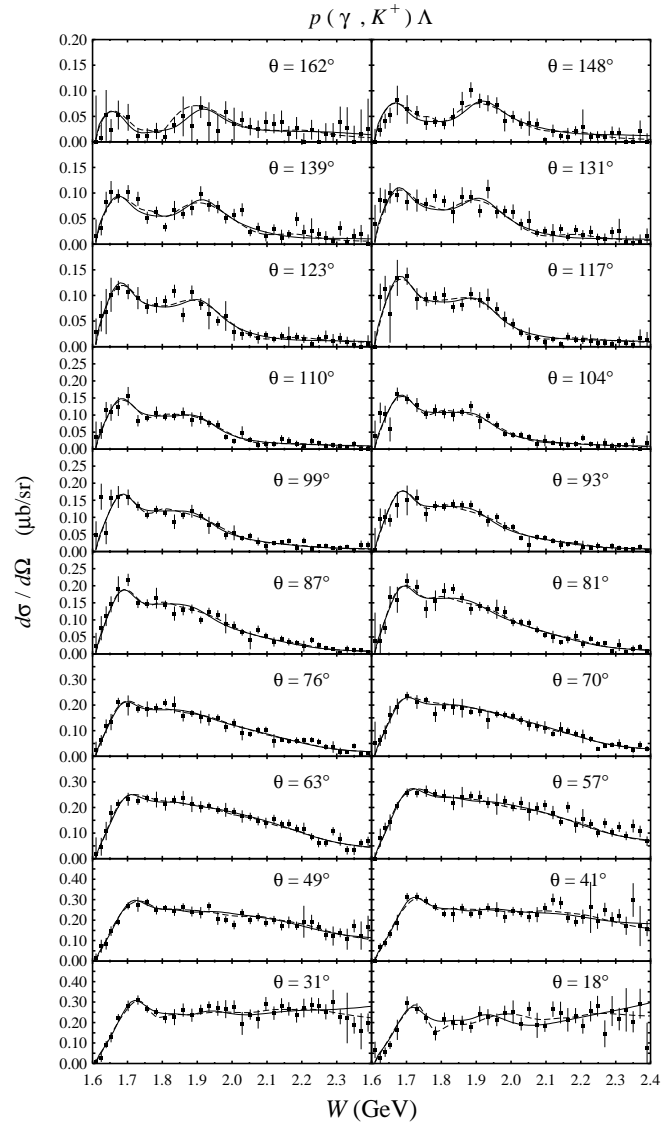}}   
\caption{Same as in Fig.\,\ref{fig:difcs}, but for $W$ distribution.
  \label{fig:difcsw}}
\end{figure}

\begin{figure}[!ht]
\centerline{\epsfxsize=9cm\epsfbox{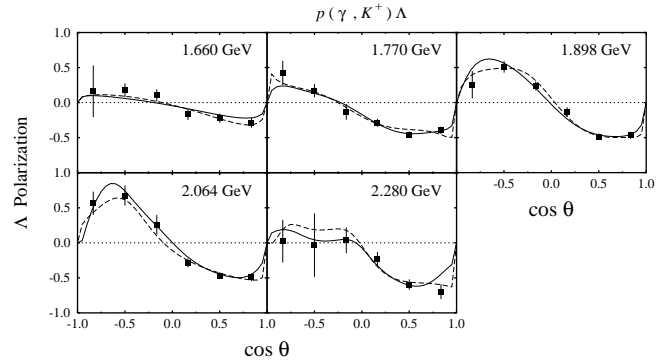}}   
\caption{The $\Lambda$ recoil polarization for the $p(\gamma,K^+)\Lambda$
  reaction. Notations are as in Fig\,\ref{fig:difcs}.
  \label{fig:pollam}}
\end{figure}

\begin{figure}[!ht]
\centerline{\epsfxsize=9cm\epsfbox{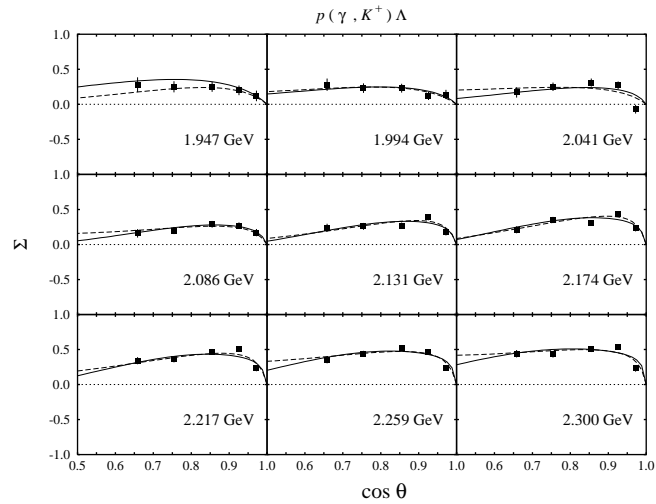}}   
\caption{Photon asymmetry for the $p(\gamma,K^+)\Lambda$
  reaction. Notations are as in Fig\,\,\ref{fig:difcs}.
  \label{fig:polph}}
\end{figure}

\begin{figure}[!ht]
\centerline{\epsfxsize=11.5cm\epsfbox{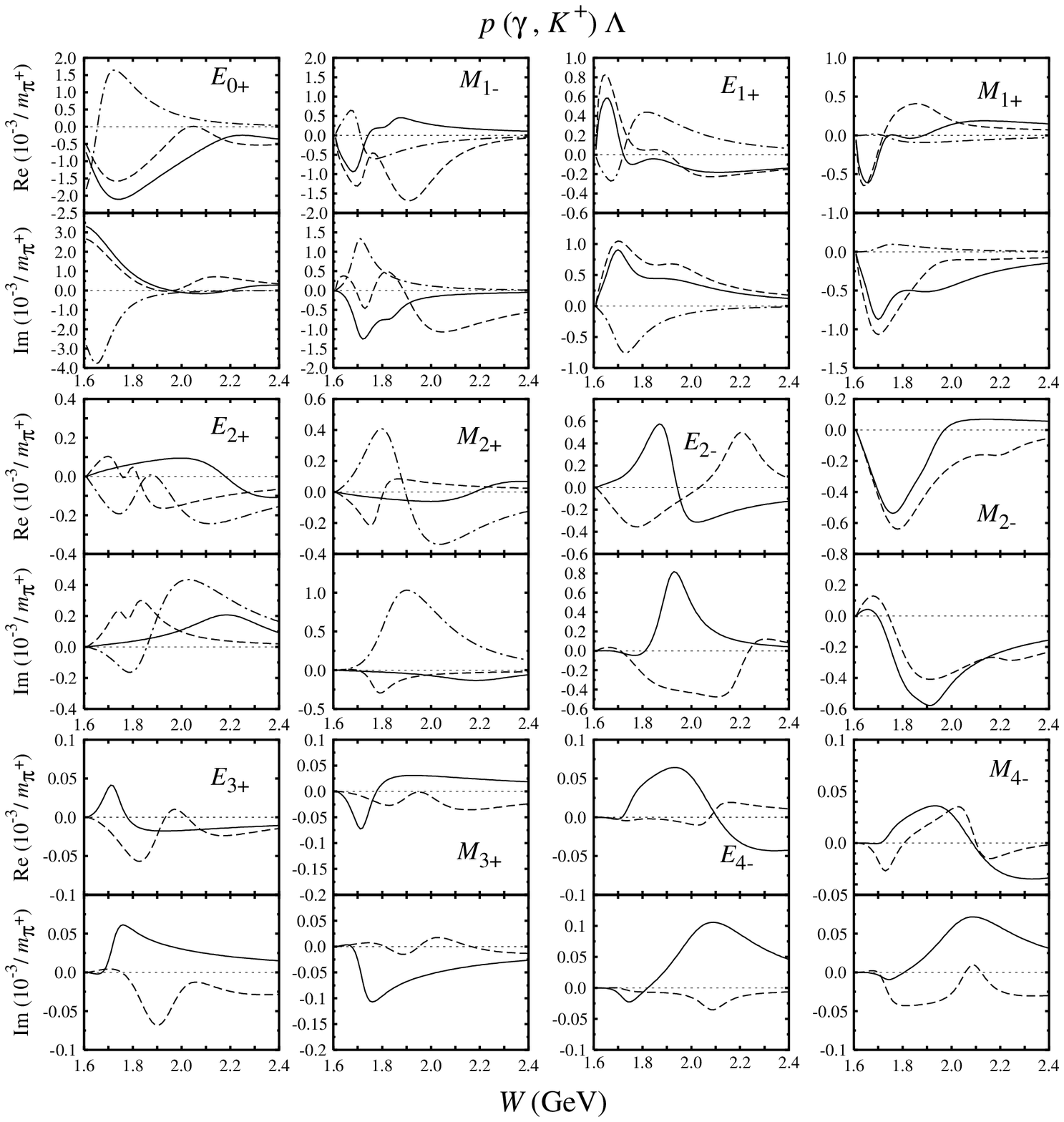}}   
\caption{Electric and magnetic multipoles for the $p(\gamma,K^+)\Lambda$
  reaction. Dashed and solid lines represent Model I and Model II
  of Table \protect\ref{tab:parameter}.
  Dash-dotted lines display the multipoles obtained from the KAON-MAID
  solution\protect\cite{kaon-maid}, where only resonances with $l\le 2$ are included.
  \label{fig:multipoles}}
\end{figure}

We summarize the results as follows:
\begin{itemize}
\item $S_{11}$ states\\
  In the two models the mass of the first $S_{11}$ does not significantly 
  change, this first $S_{11}$ is clearly related to the four-star $S_{11}$ in the PDG
  table.  A second $S_{11}$ state appears above 2 GeV but its position does not remain
fixed.
\item $P_{11}$ states\\
  The fit prefers two states in this partial wave, one at lower energy around 1700 MeV
and one in the 1850-1900 MeV mass region. The width of the lower state turns out
to be surprisingly narrow with 100 MeV, casting doubt on the interpretation that this
state is the $P_{11}(1710)$ resonance that is usually found with a much broader width
of 300 - 400 MeV. 
\item $P_{13}$ states\\
This situation is similar to the $P_{11}$ case. Two states are preferred by the fit,
 the lower with a mass below 1700 MeV, possibly corresponding to the $P_{13} (1720)$
and a state at higher energy with a broader width of 250 - 400 MeV. 
\item $D_{13}$ states\\
Using quark model predictions as guidance, a previous analysis\cite{Mart:2000ed} 
suggested that a new resonance in this partial wave, the $D_{13}(1900)$, 
provides an explanation for the new structure found in the old SAPHIR data 
around 1900 MeV.  Our new results with the new data point a more confusing and
still incomplete picture, the fits find a lower state around 1750 MeV, while the
mass of the higher state appears to be ill determined.
\item $D_{15}$ and $F_{17}$ states\\
In both partial waves, the fit clearly rejects two separate states, yet when combined
into one resonance, the mass turns out to be unstable.  It is not clear yet if resonances
are really needed in these partial waves or if these phenomena are mocking up 
missing background physics.
\item $G_{17}$ states\\
The only well-known state in this partial wave is around 2100 MeV, a new state 
with a mass as low as 1750, as suggested by our fit, would be surprising. Further
studies will have to verify if this state is in fact required to fit the data.
\end{itemize}

These sample results display the difficulty in identifying a unique set
of resonances required to fit the data.  Clearly, minimization of the $\chi^2$ 
cannot be the only criterion for establishing a new resonance but one must
achieve consistency with other reactions through a genuine multichannel analysis.

Figures 2-5 compare the model fits with the available differential cross section 
and polarization data.  As expected from the two $\chi^2$ the models are
virtually indistinguishable in these observables, yet can lead to very different 
multipoles, as shown in Figure 6. Clearly, more polarization data are needed 
that should permit a more model-independent multipole analysis.

\section*{Acknowledgment}
This work was supported in part by the QUE project (TM and AS) and the 
U.S. Department of Energy contract no. DE-FG02-95ER-40907 (CB).


\end{document}